\documentclass{aa}
\usepackage{graphics} 
\input psfig.tex 

\begin{document}

    \thesaurus{03        
              (11.01.1   
	       11.03.2	 
	       11.06.1   
	       11.19.4	 
	       11.19.5)} 

\title{ Stellar populations in	blue compact galaxies}

\author{X. Kong\inst{1}\inst{,3} 
\and F. Z. Cheng\inst{1}\inst{,2}}

\offprints{X. Kong (xkong@mail.ustc.edu.cn)\\ 
$^\dag$ BAC is jointly sponsored by the Chinese Academy of Sciences 
and Peking University.}

\institute{Center for Astrophysics, University of Science and
	      Technology of China, 230026, Hefei, P. R. China,
	\and National Astronomical Observatories,
             Chinese Academy of Sciences, 100012, Beijing, P. R. China,
	\and Beijing Astrophysics Center (BAC)$^\dag$, 100871,
	Beijing, P. R. China}

\date{Received 28 May 1999 / Accepted 29 Sept. 1999}

\maketitle

\begin{abstract}

Blue compact galaxies (BCGs) are compact galaxies that are dominated
by intense star formation.  Comparing the observational properties with
the predictions of stellar population synthesis model, we have analyzed
the nuclear stellar population and emission line spectra in a sample
of 10 BCGs.  The results indicate that the continuum flux fractions at
5870{\AA} due to old stellar components and young stellar components
are both important. The contribution from intermediate age components is
different in different galaxies. Our results suggest that BCGs are old
galaxies, in which star formation occurs in short intense burst separated
by long quiescent phases. We have also derived the internal reddening for
the stellar population by population synthesis method, and the internal
reddening for the emitting gas clouds by the Balmer line ratio. The
former is significantly smaller than the latter for BCGs. A model of
clumpy foreground dust, with different covering factors for the gas
and stars, can explain the difference. Combining the internal reddening
value and the stellar population, we have decreased the effecting from
the internal reddening and underlying stellar absorption, and accurately
measured most emission lines for each BCGs. Using these emission lines,
we have attempted to identify the ionizing mechanism of BCGs. The
ionizing mechanism for these emission line regions of BCGs is typical of
photoionization by stars, characteristics of a low extinction HII regions.

\end{abstract}

%
%

\section{Introduction}

Blue compact galaxies (BCGs) were first observed spectroscopically by
Sargent \& Searle (\cite{sargent}), who clearly established that the
properties of these galaxies implied high star formation rates at low
metallicities (Doublier et al. \cite{doublier}). BCGs have been thought
to represent a different and extreme environment for star formation
compared to the Milky Way and many other nearby galaxies. They are
very important for understanding the star formation process and galactic
evolution (Kinney et al. \cite{kinney}, Martin \cite{martin}).	BCGs are
characterized by their compact morphology and very blue UBV colors
(Sage et al. \cite{sage}, Hunter \& Thronson \cite{hunter}). Their
optical spectra show strong narrow emission lines superposed on a
nearly featureless continuum, similar to the spectrum of HII regions
(Izotov et al. \cite{izotov97}, \"Ostlin et al. \cite{ostlin}).  Radio
observations at 21-cm have shown BCGs contain large amounts of neutral
hydrogen. The mean value of $M_{\rm HI}/M_{\rm tot}$ for a sample of
122 BCGs is 0.16 (Kr\"{u}ger et al. \cite{kruger95}, Salzer \& Norton
\cite{salzer}). Systematic spectroscopic studies of BCGs  have shown that
about one-third of BCGs have broad W-R bumps, mainly at $\lambda4650$ that
are characteristic of late WN stars (Conti \cite{conti}, Izotov
et al. \cite{izotov97}).

Ever since their discovery, the question has arisen whether BCGs are truly
young systems where star formation is occurring for the first time, or
whether they are old galaxies with current starburst superposed on an old 
underlying stellar population (Garnett et al. \cite{garnett},
Lipovetsky et al.  \cite{lipovetsky}). Because the star formation rate
in BCGs is very high, the metallicity could have reached the observed value 
even within a time ${\rm T} \sim 10^8{\rm yr}$
(Fanelli et al. \cite{fanelli}). Hence, one interpretation for the low
metallicity, high gas content and high star formation rate is that BCGs
are young objects, and they are being seen at the epoch of the formation
of the first generation stars (${\rm T} \sim 10^8{\rm yr}$). The other
interpretation is that they are old objects in which star formation
occurs in short bursts with long quiescent phases in between (Kr\"{u}ger
et al. \cite{kruger95}, Gondhalekar et al. \cite{gondhalekar}, \"Ostlin
et al. \cite{ostlin}).

The low metal abundance together with the high star formation rates and
large gas masses makes BCGs most suitable to determine the element
abundance (Thuan et al. \cite{thuan95}, \cite{thuan96}), the primordial
helium abundance ${\rm Y}_p$ (Izotov et al. \cite{izotov94}) and to study
the variations of one chemical element relative to another (van Zee et
al. \cite{van98}). It also provides a wealth of diagnostics for the study 
of intrinsic physical conditions (Izotov \& Thuan \cite{izotov99}). The
results of these papers are based on direct measurements of the emission 
line intensities, but according to Vaceli et al. (1997), the observed 
emission line intensities are  affected by the underlying stellar 
absorption. Since the stellar absorption can affect substantially some 
fundamental emission lines used for the derivation of reddening and other 
physical and chemical properties, one of the first and most critical steps
in the analysis of BCG spectroscopic properties is to quantify and remove 
the contribution of the stellar population.

If we can resolve the stellar population of a BCG, we can know its age
and star formation regime. We can then subtract the stellar
absorption line from the emission-line spectrum. With the launch
of HST and 10-m class telescope, we are now witnessing a new era
that allows to analyze in detail nearby objects, such as Galactic
HII regions or 30 Dor in the LMC and resolve old red giants in a few 
distant galaxies(Grebel 1999). Such studies allow us to resolve and
study individual stars in massive star clusters. However, as one studies
objects at larger distances, individual stars (except for some giants) 
are unresolved and hence
we are limited to studying their global properties (Mas-Hesse \& Kunth
\cite{mas}). In this paper we have selected 10 BCGs and determined their 
stellar population by applying a population synthesis method based on star 
cluster integrated spectra. In a subsequent paper, we will apply an evolution 
population synthesis method to these galaxies and the results of the two 
papers will be considered together.

The outline of the paper is as follows. In Sect. 2 we describe the
observations and data reduction. In Sect.3 we present measurements
of equivalent widths and continuum analysis for the BCG spectra. In
Sect. 4 we carry out the population synthesis and give the results of
computation. In Sect. 5 we subtract the stellar population synthesis
spectra from the observed ones and study the resulting emission line
spectra. The results are presented in Sect. 6. In Sect. 7, we summarize
our conclusions. Throughout this paper, we use a Hubble constant of
${\rm H}_0=50~{\rm km ~s^{-1}~Mpc^{-1}}$.


\section{Observations and Data Reduction}

Our sample of 10 blue compact galaxies was selected from Kinney et
al. \cite{kinney}; seven of these have $M_{\rm B} > -20$ mag and
are dwarf galaxies (BCDG). Table~\ref{tab1} describes the sample
properties. The target names are listed in Column 1. Column 2 to
Column 8 respectively give the source coordinates, morphological type,
the radial velocity relative to the Local Group, photographic magnitude,
absolute magnitude and Galactic reddening.

Long-slit spectroscopic observations were carried out in March and July
of 1997.  All the observations were made with the 2.16 m telescope
at the Xinglong Station of Beijing Astronomical Observatory, using
a Zeiss universal spectrograph with a grating of 300 grooves $mm^{-1}$
dispersion. A Tek 1024$\times$1024 CCD was employed, covering a
spectral range from 3500 to 7500{\AA} with a resolution of 9.6{\AA}
(2 pixels). The slit aperture was fixed at $250 \mu m$, corresponding to
2.5$^{\prime\prime}$ projected on the sky, to match the typical seeing
value at Xinglong Station, and is set at position angle $90^{\circ}$.
All the spectra were extracted by adding the contributions
of 5 pixels around the nucleus. The sky background was estimated from a
linear interpolation of two regions located 30$^{\prime\prime}$ from the
nucleus. The last two columns of Table~\ref{tab1} list the observation
date and exposure time.

\begin{table*} 
\caption{Parameters of the selected  Blue
Compact Galaxies and Log of optical observations} 
\label{tab1}
\begin{flushleft} 
\begin{tabular}{lcccrccccc} 
\noalign{\smallskip}
\hline 

Galaxy& R.A. & Decl. &	Morph.&$v_{\rm hel,obs}$ &$B_{\rm T}$ &
$M_{\rm B}$ & $E(B-V)_G$& Obs. date & Exp. \\ 
Name& (B1950) & (B1950) &&km/s &(mag)&(mag)&&dd/mm/yy&(sec.)\\ 
\noalign{\smallskip} 
\hline
\noalign{\smallskip} 
IC1586 &00:45:17.0&22:06:07&  & 5821&14.2&-20.53&0.02&  25/07/1997& 1500 \\
NGC2537&08:09:42.5&46:08:32&Sc& 441&12.4& -18.13&0.04&  18/03/1997& 1200 \\ 
MRK19&  09:12:53.5&59:58:53& & 4230&15.6& -19.07&0.03&  18/03/1997& 1800 \\ 
MRK108& 09:17:26.1&64:26:57&I0& 1534&15.4&-17.57&0.04&  18/03/1997& 2000 \\ 
MRK25 &10:00:22.0&59:40:50&  & 2602&14.8& -19.96&0.00&  19/03/1997& 1500 \\
MRK35 & 10:42:16.5&56:13:23&Im& 995 &13.2&-18.50&0.00&  18/03/1997& 1200 \\ 
NGC4194&12:11:41.7&54:48:21&Sm& 2506&13.0&-20.71&0.00&  19/03/1997& 1500 \\ 
UGC9560&14:48:55.1&35:46:36&Ir.&1213&14.8&-17.95&0.00&  18/03/1997& 1500 \\ 
UGCA410&15:35:48.4&55:25:34&dE& 665 &15.4&-16.63&0.01&  20/03/1997& 2820 \\ 
MRK499&16:47:02.6&48:47:44& & 7710&14.9&  -21.12&0.00&  20/03/1997& 2500 \\ 
\noalign{\smallskip}
\hline 
\end{tabular} 
\end{flushleft} 
\end{table*}

The standard reduction to flux units and transformation to linear
wavelength scale were made using the IRAF\footnote{IRAF is provided
by NOAO} packages. The IRAF packages CCDRED, TWODSPEC and ONEDSPEC were
used to reduce the long-slit spectral data. The wavelength was
calibrated using a He/Ar comparison lamp. 20-odd lines were used
to establish the wavelength scale, by fitting a first-order cubic
spline. The accuracy of the wavelength calibration was better than
2{\AA}. On most nights, more than two KPNO standard stars were used for
relative flux calibration. We followed standard procedure in
the data reduction: bias subtraction, flat fielding, sky subtraction,
cosmic ray extinction, CCD response curve calibration, wavelength and
photometric calibration, and extinction correction. The dark counts
were so low that they were not subtracted. An estimate of
the radial velocity of the galaxies was made to correct to zero red
shift. The foreground reddening due to our Galaxy was corrected
using the value from Burstein \& Heiles (\cite{burstein}). Atmospheric
extinction was corrected using the mean extinction coefficients for the
Xinglong station. The final extracted spectra of the BCGs (labeled as OBS)
are shown in Figure~\ref{fig1}.


\section{Measurements}

Our main goal is to study the stellar population of BCGs. To do this,
we have used the synthesis method described in Schmitt, Bica \& Pastoriza
(\cite{schmitth}). The method minimizes the differences between the observed 
and synthetic equivalent widths for a set of spectral features. To measure
the equivalent widths accurately, it is extremely important first to have
a good fit to the continuum.  Hence the analysis of each of the
sample spectra proceeds in two steps (1) determining a pseudo-continuum
at selected pivot-points, and (2) measuring the equivalent widths (EWs)
for a set of selected spectral lines.

The continuum and EW measurements followed the method outlined in
Bica \& Alloin (\cite{bica86}), Bica (\cite{bica88}) and Bica et
al. (\cite{bica94}), Cid Fernandes et al. (\cite{cid}), and
subsequently used in several studies of both normal and emission line
galaxies (e.g., Jablonka et al. \cite{jablonka}, Storchi-Bergmann
et al. \cite{storchi}, McQuade et al. \cite{mcquade}, Bonatto et
al. \cite{bonatto}). This method first determines a pseudo-continuum
at a few pivot-wavelengths, and then integrates the flux difference
with respect to this continuum in the defined wavelength windows
(Table~\ref{tab3}) to determine the EWs. The pivot wavelengths used
in this work are based on the same as those used by the above authors; 
these were chosen to avoid regions of strong emission or absorption 
features (Table~\ref{tab2}). 
Four point flux values (3784, 3814, 3866, 3918 {\AA}) were used for the
Balmer discontinuity. The use
of a compatible set of pivot points and wavelength windows is important,
since it allows a detailed quantitative analysis of the stellar population
by the synthesis techniques using the spectral library of star clusters
(Bica \& Alloin \cite{bica86}).

The determination of the continuum was done interactively, was taking
into account the flux level, noise and small uncertainties in wavelength 
calibration, as well as the presence of emission lines. The 5870{\AA}
point, in particular, is sometimes buried underneath the HeI 5876{\AA}
emission line. In such cases, adjacent wavelength regions guided the
placement of the continuum.  The final measured fluxes, normalized to
the flux at 5870{\AA}, are given in Table~\ref{tab2}.  After fitting the
continuum points to a continuum spectrum, we measured the equivalent
widths of seven characteristic absorption lines. When a noise spike
was present in the wavelength window, it was
necessary to make `cosmetic' corrections. The results are given
in Table~\ref{tab3}.  The first two rows give the names of the spectral
lines and corresponding wavelength range. The EWs are in units of {\AA},
and negative sign denotes emission lines.

\begin{table*} 
\caption[]{Continuum points $F_{\lambda}$ relative to
$F_{\lambda}$(5870{\AA}), corrected for the foreground reddening}
\begin{flushleft} 
\label{tab2} 
\begin{tabular}{lcccccccccc}
\noalign{\smallskip} 
\hline 
\noalign{\smallskip} 
Name & $F_{3660}$ & $F_{3784}$ & $F_{3814}$ & $F_{3866}$ &
$F_{3918}$ & $F_{4020}$ & $F_{4570}$ & $F_{4630}$ & $F_{5313}$
& $F_{6630}$ \\ 
\noalign{\smallskip} 
\hline 
\noalign{\smallskip}
IC1586&1.025&1.193&0.996&1.268&1.399&1.311&1.206&1.231&1.045&0.959 \\
NGC2537&0.622&0.563&0.739&0.686&0.935&0.891&0.985&1.023&0.907&0.796 \\
MRK19  &1.255&1.189&1.933&2.758&2.201&1.790&1.314&1.249&0.989&0.866 \\
MRK108 &1.108&0.868&2.224&2.207&1.840&1.895&1.041&1.039&0.847&0.886 \\
MRK25  &1.100&1.113&1.215&1.394&1.378&1.338&1.206&1.206&1.089&0.870 \\
MRK35  &1.365&1.078&1.244&1.445&1.559&1.776&1.346&1.338&1.114&0.865 \\
NGC4194&0.764&0.776&0.921&1.154&0.957&1.026&0.900&0.922&0.962&0.964 \\
UGC9560&1.277&1.538&1.664&1.901&1.843&1.613&1.417&1.295&0.975&0.829 \\
UGCA410&0.827&1.422&1.588&1.798&1.540&1.269&1.153&1.114&1.026&0.966 \\
MRK499&1.012&1.331&1.216&1.442&1.551&1.440&1.252&1.252&1.033&0.935 \\
\noalign{\smallskip} 
\hline 
\end{tabular} 
\end{flushleft} 
\end{table*}

\begin{table*} 
\caption[]{Measured equivalent widths in {\AA} for blue compact galaxies.} 
\label{tab3} 
\begin{flushleft}
\begin{tabular}{lcccccccc} 
\noalign{\smallskip} 
\hline
\noalign{\smallskip} 
Galaxy& KCaII  &${\rm H}_{\delta}$ &    CN &$G_{band}$ &${\rm H}_{\gamma}$ &${\rm H}_{\beta}$ &MgI+MgH \\ 
Name &3908-3952&4082-4124&4150-4214&4284-4318&4318-4364&4846-4884&5156-5196 \\ 
\noalign{\smallskip} 
\hline 
\noalign{\smallskip}
IC1586   &5.845 &0.287  &-2.73  &2.122 &-0.89 &-6.90 &1.029\\
NGC2537  &18.96 &2.293  &5.692  &3.329 &-14.7 &-33.9 &4.270\\ 
MRK19    &9.878 &5.313  &-0.37  &3.786 &-4.21 &-35.5 &4.138\\ 
MRK108   &18.25 &-8.92  &8.427  &4.192 &-37.1 &-105. &3.434\\ 
MRK25    &4.318 &3.482  &3.954  &2.682 &-2.39 &-8.96 &2.895\\ 
MRK35    &2.769 &-6.29  &7.278  &1.936 &-15.1 &-22.6 &3.375\\
NGC4194  &3.707 &1.261  &0.449  &1.919 &-2.62 &-10.1 &3.226\\ 
UGC9560  &4.736 &-4.67  &4.230  &3.678 &-10.7 &-40.5 &4.303\\ 
UGCA410  &10.61 &1.073  &14.33  &8.688 &-13.0 &-50.6 &3.237\\ 
MRK499   &3.687 &3.092  &-1.07  &0.960 &4.124 &-1.11 &1.161\\ 
\noalign{\smallskip}
\hline 
\end{tabular} 
\end{flushleft} 
\end{table*}

\section{Stellar Population Synthesis}

To study the history of star formation, age, and ionization mechanism of
BCGs, we applied the synthesis method of Schmitt et al. (\cite{schmitth})
for a determination of the stellar population in their nuclear region.  
In this method, we start with a sample of star clusters, consists of 3
clusters in the SMC, 12 in the LMC, 41 Galactic globular clusters and
3 rich compact Galactic open cluster, together with 4 HII regions
(Bica \& Alloin \cite{bica86}). Then a grid of base components 
(comprising the values of the continuum at selected points and of the 
EWs of selected absorption lines) is constructed by interpolation and 
 extrapolation in the ${\rm Age} - [Z/Z_{\odot}]$ plane of those 
quantities of the clusters of the sample.
The grid consists of 34 points with ages at $10^7, 5\times10^7, 10^8, 
5\times10^8, 10^9, 5\times10^9, {\rm and} >10^{10} {\rm yr}$, and
 $[Z/Z_{\odot}]$= 0.6, 0.3, 0.0, -0.5, -1.0,-1.5, -2.0, plus one point
representing the HII region. The method, when applied to a given target,
consists of adjusting the percentage contributions of the 35 base 
components by minimizing the difference between the resulting 
and measured EWs of the selected set of absorption lines. When all
the resulting EWs reproduce those of the galaxy within allowed limits, 
we said to have obtained an acceptable solution. We then take all the 
acceptable solutions to form an average solution (Schmitt et al. 
\cite{schmitth}).

The computation can be performed in two ways: one way spans the whole
${\rm Age} - [Z/Z_{\odot}]$ plane (multi-minimization procedure,
hereafter MMP), while the other is restricted to chemical evolutionary
paths through the plane (direct combination procedure, hereafter DCP). We
combine these two methods in our paper.  We first use the MMP method to
single out the main contributing components. And then, based on their
resemblance to the whole-plane solution and on their reduced chi-square
($\chi^2$), we select the best evolutionary path and use the DCP method
to give the final result.

\subsection{The results of MMP}

A detailed analysis of the MMP method can be found in Schmidt et
al. (\cite{schmidta}). 
This method searches the vector space of resolutions generated
by the entire 35 component basis, leading to a representative set of
acceptable solutions to the synthesis problem. We tried various combinations
of the 35 components until a good match between the equivalent widths of 
the synthetic and observed lines is obtained. An iterative optimization 
procedure was used, and each iteration alters the percentages of different 
components. The input parameters are the measured equivalent widths of the 
selected absorption lines, the continuum ratios and a set of trial values of 
$E(B-V)$ between 0.0 and 1.0 at steps of 0.02. 
The result is expressed with the flux fractions at 5870{\AA} for each component.
The flux fractions of different components for the 10 BCGs are shown in
Table~\ref{tab4} (HIIR denotes the HII region component).

\begin{table*} 
\caption{Flux fractions (\%) at 5870{\AA} for the whole plane solution for
 BCGs.}
\label{tab4} 
\begin{flushleft} \begin{tabular}{llllllll|r|llllllll}
\hline 
HIIR& E7 & 5E7 & E8 &  5E8 &  E9 &  5E9 & $>$ E10 & [$Z/Z_\odot$]&HIIR & E7 & 5E7 & E8 &  5E8 &  E9 &  5E9 & $>$ E10  \\ 
\hline 
\multicolumn{8}{c|}{a) IC1586}&&
\multicolumn{8}{|c} {b) NGC2537 }\\ 
\cline{1-8}\cline{10-17}
&2.37&0.45&0.45&0.00&0.00&0.00&0.00&0.6&&1.00&0.06 &1.36 &2.99&2.52 &1.23 &0.04\\
&1.05&0.45&0.45&0.00&0.00&0.00&0.00&0.3&&0.90&0.92&1.16&3.81&3.65&1.06&0.13\\
&2.30&0.45&0.45&0.00&0.00&0.00&0.00&0.0&0.00&0.75&5.79&5.02&10.1&11.5&6.34&1.11\\
3.36&6.43&2.08&2.04&14.6&0.00&0.00&0.00&-0.5&&0.62&0.16&0.40&2.22& 3.04&2.53 & 3.33\\ 
&&&&&0.76&0.61&0.61&-1.0&&&&&&4.96&5.46&4.39\\
&&&&&&4.72&6.22&-1.5&&&&&&&4.61&4.57\\ 
&&&&&&&50.1&-2.0&&&&&&&&3.24\\
\noalign{\smallskip} 
\multicolumn{8}{c|}{c) MRK19}&&
\multicolumn{8}{|c} {d) MRK108 }\\ 
\cline{1-8}\cline{10-17} &0.10&1.84 &2.33&8.63&1.29 &0.00&0.00&0.6&&0.00 &0.14&0.21&13.5&1.84&0.28&0.00\\ 
&0.10 &1.08&3.54&10.3&1.74&0.00&0.00&0.3&&0.00&0.00&0.07&14.1&1.90&1.21&0.00\\
0.00&0.16&4.78&5.97&26.8&6.49&0.13&0.26&0.0&0.00&0.00&0.14&0.28&31.7&9.10&2.13&0.70\\
&0.16&1.78&2.61&0.13&1.82&0.43&0.40&-0.5&&0.00&0.00&0.14&0.00&1.69&1.56&0.78\\
&&&&&3.34&1.19 &1.33&-1.0&&&&&&4.18&2.27&4.68\\
&&&&&&3.11&3.94&-1.5&&&&&&&3.69&4.68\\ &&&&&&&4.17&-2.0&&&&&&&&2.63\\
\noalign{\smallskip} 
\multicolumn{8}{c|}{e) MRK25}&&
\multicolumn{8}{|c} {f) MRK35 }\\ 
\cline{1-8}\cline{10-17}
&0.16&0.68&1.05&3.74&2.48&0.67&0.02&0.6&&0.34&3.33&3.28&3.98&0.27&0.00&0.00\\
&0.17&1.76&1.10&4.68&2.14&1.09&0.34&0.3&&1.47&3.10&2.72&3.84&1.99&0.07&0.16\\
23.7&0.20&2.74&7.11&10.8&7.00&1.71&0.85&0.0&22.2&4.93&11.0&10.6&12.1&2.59&0.97&0.20\\
&0.24&1.66&2.85&0.53&1.31&1.64&1.63&-0.5&&1.23&2.20&3.87&0.00&0.03&0.00&0.00\\
&&&&&2.88&1.77&1.93&-1.0&&&&&&0.30&0.00&0.00\\
&&&&&&2.24&3.20&-1.5&&&&&&&0.57&0.39\\ 
&&&&&&&3.94&-2.0&&&&&&&&0.31\\
\noalign{\smallskip} 
\multicolumn{8}{c|} {g) NGC4194}&&
\multicolumn{8}{|c} {h) UGC9560} \\ 
\cline{1-8}\cline{10-17}
&4.50&2.63&2.64&0.12&0.02&0.03&0.30&0.6&&1.90&1.03&0.72&1.82&1.66&1.25&0.27\\
&4.79&3.29&3.40&0.34&0.12&0.11&0.50&0.3&&1.23&1.27&1.98&1.73&1.73&2.02&0.27\\
&5.20&4.04&4.04&0.80&0.35&0.33&0.84&0.0&28.2&4.87&6.35&3.84&6.72&9.99&7.77&0.31\\
10.3&6.02&5.75&5.73&3.71&1.68&1.61&1.96&-0.5&&1.84&1.02&1.28&0.57&2.78&1.59&0.8\\
&&&&&3.17&3.02&3.07&-1.0&&&&&&0.94&0.76&0.71\\
&&&&&&4.59&4.40&-1.5&&&&&&&0.23&0.25\\ 
&&&&&& &6.42&-2.0&&&&&&&&0.16\\
\noalign{\smallskip} 
\multicolumn{8}{c|} {i) UGCA410}&&
\multicolumn{8}{|c} {j) MRK499} \\ 
\cline{1-8}\cline{10-17}
&0.45&0.70&0.35&0.00&3.20&0.00&0.00&0.6&&2.06&1.69&1.67&0.00&0.00&0.00&0.00\\
&1.68&0.44&0.00&4.08&1.15&0.00& 0.00&0.3&&2.47&2.07&1.75&0.00&0.00&0.00&0.00\\
0.45&9.39&2.47&0.03&21.9&10.1&1.36&0.00&0.0&&4.21&5.15&5.04&0.00&0.00&0.00&0.00\\
&3.85&0.62&0.03&1.24&4.45&0.78&0.48&-0.5&6.30&10.3&8.53&8.25&1.62&0.00&0.00 &0.00\\ 
&&&&&8.33&2.52&1.78&-1.0&&&&&&0.00&0.00&0.00\\
&&&&&&4.45&4.54&-1.5&&&&&&&1.95&3.04\\ 
&&&&&&&8.18&-2.0&&&&&&&&33.9\\
\hline 
\end{tabular} 
\end{flushleft} 
\end{table*}

From the results of the MMP method, we find some obvious trends for
all the BCGs. First, the dominant population are young (${\rm T}
\leq 5\times10^8{\rm yr}$) stellar clusters.  Second, there is a small
population of old (${\rm T}>10^{10}{\rm yr}$) and high metallicity
($[Z/Z_{\odot}]\geq0$) globular clusters.  Within old globular clusters
low metallicity components contribute more than high metallicity components.
Third, for the young and intermediate age 
(${\rm T}\sim10^9-5\times10^9{\rm yr}$) clusters, components with metallicities 
below or equal to the solar value make a large contribution. It shows that 
the stars in the BCGs have low metallicities,
and this is consistent with our previous knowledge. Lastly,
the younger components show a large dispersion in the plane, with no
clear evolutionary paths. This could be due to the fact we have only
used spectral data in the visible range. Additional data in the near
ultraviolet and infrared will produce better-constrained solutions in
the plane. Nevertheless, the presence of starbursts with $T \approx
5\times10^8{\rm yr}$ is easily recognized from the MMP results.

\subsection{The results of DCP}

To reduce the dispersion in metallicity, all the population synthesis
so far made assumed some arbitrarily chosen chemical evolution path in
the ${\rm Age} - [Z/Z_{\odot}]$ plane. The improvement in the method of
this paper is: we shall pick out from the MMP results those components
that contribute most importantly and use them to define paths of
chemical evolution, thus reducing the degree of arbitrariness. 
The 3 bright BCGs appear to follow a path containing the 11 components 
along the time sequence ${\rm T}=10^7\sim5\times10^9{\rm
yr}$ at fixed metallicity $[Z/Z_{\odot}]=-0.5$, the metallicity sequence
$[Z/Z_{\odot}]=-0.5 \sim -2.0$ with ${\rm
T}>10^{10}{\rm yr}$, and the HII region. The 7 BCDGs follow a path containing
the 12 components along the time sequence ${\rm T}=10^7\sim5\times10^9{\rm
yr}$ at fixed metallicity $[Z/Z_{\odot}]=0.0$,	the metallicity sequence
$[Z/Z_{\odot}]=0.0 \sim -2.0$ with ${\rm T}>10^{10}{\rm yr}$, and the HII 
region. In addition, we also tested other possible paths with different maximum 
metallicities; we found that the $\chi^2$ of the path selected from MMP is 
the smallest.

Table~\ref{tab5} reports the path solution. The numbers of acceptable solutions 
and corresponding reduced chi-squared ($\chi^2$) are also given. There is a 
great similarity between the path (DCP) and the whole plane (MMP) solution.
Table 5a, 5g, 5j provide the results
for the 3 bright BCGs (IC1586, NGC4194 and MRK499). We see that the
younger components with age $10^7-10^8${\rm yr} make an appreciable
contribution, $\geq35\%$. The old globular clusters make even larger
contributions. The intermediate-age components are small.  The other
tables in Table~\ref{tab5} provide the results for the 7 BCDGs.  The
dominant stellar components ($>30\%$) are in the young age bins (${\rm
T}\leq5\times10^8{\rm yr}$). The old globular clusters have different
values in  different galaxies; in some, lower than $10\%$,
and in others as much as $35\%$. The BCDGs differ from the bright BCGs in
two obvious aspects. First, the intermediate age components in the BCDGs amount to
$20\%$, higher than in the bright BCGs. A peak occurs at about $5\times10^8
{\rm yr}$, which indicates that an enhanced star formation event
occurred at that epoch. Second, the youngest components (${\rm T}=10^7{\rm
yr}$) in the BCDGs are much lower than in the bright BCGs; probably indicating
that the star formation rate of BCDGs is lower now. The HII region component
is a featureless continuum, which acts in the synthesis as a dilutor of
absorption lines. From Table~\ref{tab5}, we note that some BCDGs have
large contributions from the young components indicating 
intense star formation, and small contribution from HII regions, which
suggests that intense starbursts  have converted most gas into stars.

\begin{table*} 
\caption{Flux fractions (\%) at 5870{\AA} for the best path solution of BCGs.}
\label{tab5} 
\begin{flushleft} 
\begin{tabular}{llllllll|r|llllllll}
\hline 
 HIIR & E7 & 5E7 & E8 &  5E8 &	E9 &  5E9 & $>$ E10 & [$Z/Z_\odot$]&HIIR &
 E7 & 5E7 & E8 &  5E8 &  E9 &  5E9 & $>$ E10  \\ 
\hline
\multicolumn{8}{c|}{a) IC1586 :  Acceptable Solutions: 78, $\chi^2=1.23$}&& 
\multicolumn{8}{|c}{b) NGC2537:  Acceptable Solutions: 75, $\chi^2=1.08$ }\\ 
\cline{1-8}\cline{10-17}
&&&&&&&&				 0.6&&&&&&&&\\
&&&&&&&&				 0.3&&&&&&&&\\ 
&&&&&&&& 0.0&0.02&2.73&5.75&6.29&26.3&15.2&9.54&4.37\\
5.69&7.57&4.72&4.81&18.8&0.52&0.41&0.48&-0.5&&&&&&&&7.62\\
&&&&&&&				 2.28 & -1.0&&&&&&&&8.31\\
&&&&&&&				 7.62 & -1.5&&&&&&&&7.87\\
&&&&&&&				 47.1 & -2.0&&&&&&&&5.99\\
\noalign{\smallskip} 
\multicolumn{8}{c|}{c) MRK19 :  Acceptable Solutions: 143, $\chi^2=1.30$}&& 
\multicolumn{8}{|c}{d) MRK108:  Acceptable Solutions:  93, $\chi^2=0.87$ }\\ 
\cline{1-8}\cline{10-17} 
&&&&&&&& 0.6&&&&&&&&\\ 
&&&&&&&&				       0.3&&&&&&&&\\
0.25&1.81&8.93&12.2&52.4&10.7&4.24&2.67&0.0&0.00&0.86&3.66&3.66&44.1&32.4&10.0&2.47\\
&&&&&&&				 1.44&-0.5&&&&&&&&1.40\\
&&&&&&&				 2.47&-1.0&&&&&&&&1.08\\
&&&&&&&				 0.66&-1.5&&&&&&&&0.21\\
&&&&&&&				 2.22&-2.0&&&&&&&&0.21\\
\noalign{\smallskip} 
\multicolumn{8}{c|}{e) MRK25:  Acceptable Solutions: 133, $\chi^2=1.12$}&& 
\multicolumn{8}{|c}{f) MRK35:  Acceptable Solutions: 106, $\chi^2=0.73$}\\ 
\cline{1-8}\cline{10-17} 
&&&&&&&& 0.6&&&&&&&&\\ 
&&&&&&&&				       0.3&&&&&&&&\\
22.5&2.10&7.17&9.40&19.8&12.1&6.70&3.87&0.0&16.5&8.57&21.9&21.1&18.6&7.49&2.80&0.61\\
&&&&&&&				 3.88 &-0.5&&&&&&&&0.73\\
&&&&&&&				 3.75&-1.0&&&&&&&&0.94\\
&&&&&&&				 4.33&-1.5&&&&&&&&0.47 \\
&&&&&&&				 4.38&-2.0&&&&&&&&0.41\\
\noalign{\smallskip} 
\multicolumn{8}{c|} {g) NGC4194:  Acceptable Solutions: 54, $\chi^2=1.03$}&& 
\multicolumn{8}{|c} {h) UGC9560:  Acceptable Solutions: 52, $\chi^2=0.55$} \\ 
\cline{1-8}\cline{10-17}
&&&&&&&&				 0.6&&&&&&&&\\
&&&&&&&&				 0.3&&&&&&&&\\ 
&&&&&&&& 0.0&29.5&4.28&8.01&7.26&15.0&14.7&11.4&3.69\\
18.5&7.90&5.53&5.47&20.2&2.36&2.72&4.39&-0.5&&&&&&&&2.47\\ 
&&&&&&& 8.63 & -1.0&&&&&&&&1.72\\ 
&&&&&&&			   11.0 & -1.5&&&&&&&&1.16\\ 
&&&&&&&			    13.3 & -2.0&&&&&&&& 0.73\\ 
\noalign{\smallskip} 
\multicolumn{8}{c|}{i) UGCA410:   Acceptable Solutions:  95, $\chi^2=0.86$}&&
\multicolumn{8}{|c}{j) MRK499 :   Acceptable Solutions: 151, $\chi^2=0.53$} \\ 
\cline{1-8}\cline{10-17} 
&&&&&&&& 0.6&&&&&&&&\\ 
&&&&&&&&					0.3&&&&&&&&\\ 
9.99&0.66&2.63&2.30&28.5&19.3&6.16&0.93&  0.0&&&&&&&&\\ 
&&&&&&& 2.55& -0.5&5.28&17.3&19.9&19.3&3.70&0.51&0.30&0.31\\ 
&&&&&&& 4.81& -1.0&&&&&&&&1.41 \\ 
&&&&&&&			     9.92& -1.5&&&&&&&&5.19 \\ 
&&&&&&&			       12.2& -2.0&&&&&&&&26.8 \\ 
\hline 
\end{tabular}
\end{flushleft} 
\end{table*}


\section{Synthesized Spectra}

\begin{figure*} 
\resizebox{15cm}{!}{\includegraphics{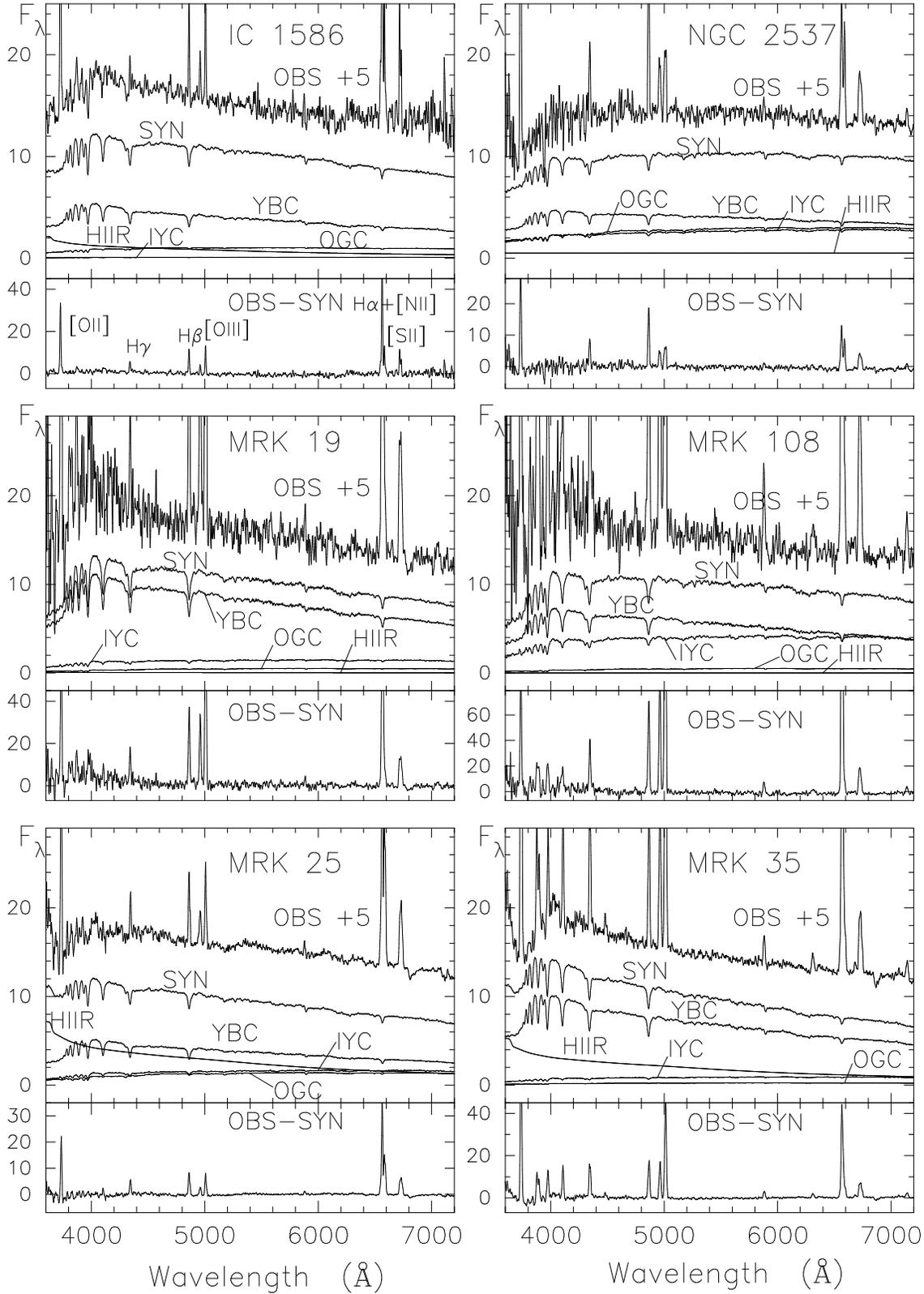}}
\caption{Observed spectra for our 10 BCGs
are shown together with the best-match population synthesis models
(cf. Table~\ref{tab5}).  The observed galaxy spectrum (OBS) has been
corrected for the foreground reddening E(B-V)$_G$ and shifted by 5 units.
In the synthetic spectrum (SYN), the young population (YBC) dominates the
light, but the intermediate age component (IYC) is also important. The emission
line spectrum appears in the (OBS-SYN) difference, in the lower part
of the figure. OBS and SYN are normalized to ${\rm F}_{\lambda} =10 $
at 5870{\AA}.} 
\label{fig1} 
\end{figure*}

We now display the results of stellar population synthesis in a more
visible form in Figure~\ref{fig1}. OBS represents the observed spectrum
of galaxy (OBS+5, the +5 indicating that it is displaced upwards by
5 units). SYN represents the synthetic spectrum resulting from the 
percentage contribution in Table 5.
The various stellar components are designated
as follow. OGC stands for the contribution from the old globular
cluster (${\rm T}>10^{10}{\rm yr}$); IYC for the contribution from the
intermediate age star cluster (${\rm T}\sim10^9-5\times10^9{\rm yr}$); YBC
for the contribution from the young blue star clusters $10^7 \leq {\rm T}
\leq 5 \times10^8{\rm yr}$), and HIIR stands for the contribution from
HII regions.  The emission line spectrum (OBS-SYN) resulting from 
subtracting SYN from OBS, is shown in the lower part of figure.

The figure shows that the synthesized spectrum gives a good fit to the
observed continuum and absorption lines for each galaxy. We can conclude
that the continuum of BCGs comes both from the stars (particularly the
young and intermediate-age stars) and HII regions. It may be interpreted
that the main energy sources of BCGs are young hot O, B stars, which
lead to the formation of HII regions around them.

The stellar subtracted spectra (OBS-SYN) can be used to study the
emission lines. We have measured the main strong emission line intensities
with Gaussian fits. The results, relative to that of ${\rm H}\alpha$
are shown in Table~\ref{tab6}.	At this stage our spectra are corrected
for the foreground reddening and the internal reddening from the
stellar populations (see next section). First, we have used the ${\rm
H}\alpha/{\rm H}\beta$ ratio in Table~\ref{tab6} to derive the internal
reddening value associated with the line-emitting regions. It will be
calculated in next section. Second, we have attempted to identify the
ionizing mechanism in these nuclei, using the emission line ratios in
the visible region.

We compared emission-line ratios calculated from Table~\ref{tab6}
with the diagnostic diagram ${\rm [NII]}6584/{\rm H}\alpha - {\rm
[OII]}3727/{\rm[OIII]}5007$ from Baldwin, Phillips \& Terlevich
(\cite{baldwin}). We plot the results in Figure~\ref{fig2}. From this
figure, we find that the BCGs are always located near or within the HII
region.  None of them are located in the loci
of planetary nebulae, power-law, or shock-heated region. This result
indicates that the young, massive mass stars formed in the nucleus are
heating the gas in the nucleus of BCGs.  This result is very similar to
those found from population synthesis. We also show in this diagram the
effect of internal reddening.  We can see that it, even in the case of 
NGC4194, with the strongest reddening derived from its line-emitting 
regions, is not large.


\section{Results }

From the stellar population analysis in Sect. 4 and the emission line
spectrum in Sect. 5, combined with results from other studies, it is
now possible to reveal some global properties of BCGs.

\subsection{Age and Star Formation Rate}

There are two major competing theories for BCGs. The first one claims
that BCGs are truly young systems undergoing the first star formation
episode in the galaxy's lifetime. The second model suggests that BCGs are
old galaxies, which are mainly composed of older stellar populations, 
with, however, a brief episode of violent star formation, in 
order to account for the observed spectroscopic features and spectral 
energy distributions.

Using the population synthesis method, we know the stellar population
(different ages and metallicities) percentage contributions at
$\lambda=5870${\AA} for each of the 10 BCGs. These results show clearly that,
for each, the old globular clusters and intermediate age components
(${\rm T}\leq10^9 {\rm yr}$) make sizeable contributions to the galactic
spectrum. The presence of large fractions of old or intermediate age
components indicates that the star formation happened already at an
early stage, and at a high rate. Our
results support the second model that BCGs are old galaxies.

For IC1586, NGC4194 and MRK499, the contributions coming from the young
and the old stellar components are large, but that from the intermediate age
component is small. It suggests that the rate of star formation during the
intermediate age period is smaller than in the other periods, and the star
formation process is not continuous in these galaxies. For the 
BCDGs, the contribution from the intermediate age component
is important, star formation was most vigorous in its intermediate age
period, with relatively small contributions from the other periods.
This also implies that star formation in these galaxies is also discontinuous.

The other result of our population synthesis is that while the observed
properties of the bright BCGs (IC1586, NGC4194, MRK499) and the BCDGs are very
similar, their stellar components and star formation regimes are
generally different. There are many old and young stellar components
in bright BCGs, and its recent star formation rate is very high. For
BCDGs, the old and young stellar components are relatively small, but
the contribution from intermediate age stellar populations is important.

The stellar population synthesis suggests that BCGs are old galaxies,
in which the process of star formation is intermittent. Star formation
has been violent in one of its evolution periods.  These results are
also supported by other observations (Papaderos et al. \cite{papaderos},
Sung et al. \cite{sung}, Aloisi et al. 1999). 
It illustrates that the present method is more
than a simple population synthesis since it provides a direct estimate
of the chemical evolution of the galaxy.

\begin{figure}
\psfig{figure=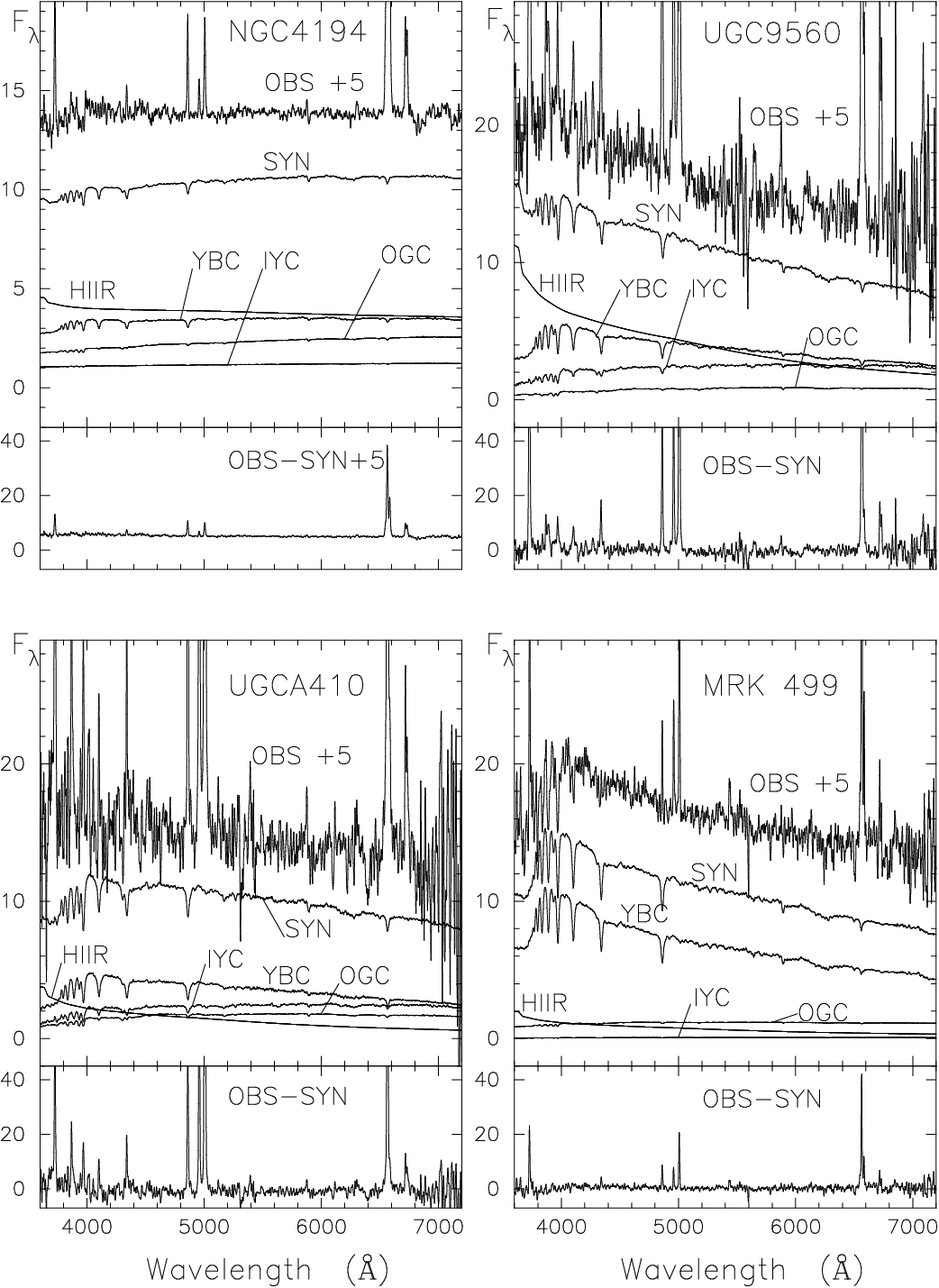,width=8.8cm,height=10.5cm}   
\begin{center}
Fig.1 continue.
\end{center}  
\label{Fig.1} 
\end{figure}


\begin{table*} 
\caption{Emission line intensities with respect to ${\rm H}\alpha$ 
(normalized to 100) and the internal reddening value of BCGs.} 
\label{tab6} 
\begin{flushleft} 
\begin{tabular}{lrrrrrrrrrrrr}
\noalign{\smallskip} 
\hline 
\noalign{\smallskip} 
Galaxy&[OII]&${\rm
H}\gamma$&${\rm H}\beta$&[OIII]&[OIII]&[NII]&[SII]&
$X^*$& $Y^{**}$& 
\multicolumn{3}{c}{E(B-V)}\\ 
\cline{11-13}
Name&3727&4340&4861&4959&5007&6584&6717,31&&&${\rm H}\alpha/{\rm
H}\beta$&MMP&DCP\\ 
\hline 
IC1586&  65.62& 11.29& 20.15& 10.00& 23.22& 16.83& 36.47&  0.45& -0.77&  0.51&0.13&0.14 \\ 
NGC2537&185.29& 45.84&104.32& 34.37& 56.40& 59.05& 66.29&  0.52& -0.23&      &0.13&0.12 \\ 
MRK19&   52.61& 11.95& 28.09& 24.12& 78.23&	7.08& 23.18& -0.17& -1.15&	0.20&0.03&0.03 \\ 
MRK108&  26.19& 11.87& 24.18& 32.45&104.44& 2.09& 11.22& -0.60& -1.68&  0.34&0.03&0.03 \\ 
MRK25&   71.44& 16.36& 29.08& 12.85& 25.32& 59.58& 37.77&  0.45& -0.22&  0.17&0.06&0.04 \\
MRK35 & 141.15& 31.25& 37.82& 37.81& 99.38& 24.01& 23.55&  0.15& -0.62&	    &0.05&0.04 \\ 
NGC4194& 20.77&  5.24& 15.63&  4.73& 14.58& 48.59& 24.84&  0.15&  0.31&  0.75&0.33&0.35 \\ 
UGC9560&  67.07&  9.32& 26.21&	3.19&108.70& 10.84& 15.68& -0.21& -0.96&  0.27&0.03&0.02 \\
UGCA410& 28.25&  9.15& 26.77& 47.21&159.02&  7.61& 13.01& -0.75& -1.12& 0.25&0.02&0.02 \\ 
MRK499& 54.83&  8.91& 21.23& 19.84& 49.43& 25.50& 20.00& 0.04& -0.59& 0.46&0.15&0.16\\ 
\noalign{\smallskip} 
\hline 
\end{tabular}
\end{flushleft} 
$^*X \equiv \log({\rm [OII]}3727/{\rm [OIII]}5007)$;
$^{**}Y \equiv \log({\rm [NII]}6584/{\rm H}\alpha)$.  
\end{table*}

\begin{figure} 
\psfig{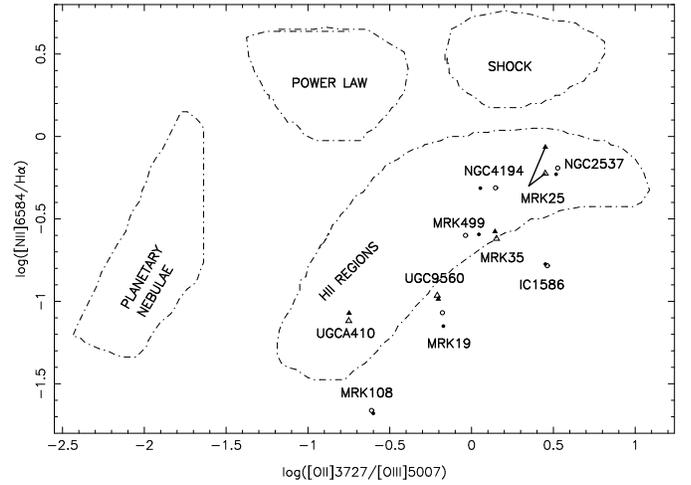}
\caption{ Logarithmic
plot of the line intensity ratios {\rm [NII]}$\lambda6584/{\rm H}\alpha$
versus {\rm [OII]}$\lambda3727$/{\rm [OIII]}$\lambda5007$.  Open symbols
correspond to data corrected for the foreground reddening.  Black symbols
represent data points with an additional reddening correction intrinsic
to the line emitting regions.} 
\label{fig2} 
\end{figure}
                                                     
\subsection{Internal Reddening}

When we investigate the internal energy source, physical condition
and internal structure of galaxies, we must take into account the
effect of internal reddening (Pizagno \& Rix \cite{pizagno}). The
effect of dust extinction on the emerging radiation is one of the
least understood physical phenomena (Calzetti \cite{calzetti97}, Ho
et al. \cite{ho}). To study the internal reddening properties of BCGs,
we quantify the discrepancy between the dust extinction measured from
the emission line ratios and the optical continuum.

In the method of population synthesis used in this paper, the internal
reddening is taken as an adjustable parameter, so that an estimate for
the internal reddening is made at the same time as the stellar composition.
We try various values of the internal reddening, make the appropriate
correction in the continuum spectrum, then use this corrected continuum
spectrum in the synthesis to find the best solution.  This is an
empirical way of determining the galactic reddening, its advantage is that it is
assumption-free. The values of galactic internal reddening are listed in
Table~\ref{tab6}, $E(B-V)_{\rm MMP}$ is the result from the MMP method,
$E(B-V)_{\rm DCP}$ is the result from the DCP method.  We find that the
values are small ($E(B-V)\leq0.35$), which is consistent with the BCGs
being metal-poor and dust-poor. We find the reddening also clearly
depends on the shape of the spectrum. The flattest spectrum
(NGC4194) goes with the largest color excess. The steeper spectrum,
the less the extinction.

The Balmer line ratio ${\rm H}{\alpha}/{\rm H}{\beta}$ allows us to
characterize the dust extinction in the regions where the nebular lines
are produced. We measured the internal reddening value $E(B-V)_{{\rm
H}{\alpha}/{\rm H}{\beta}}$ of the 10 BCGs using the observed emission lines
${\rm H}{\alpha}$ and ${\rm H}{\beta}$. The difference in the calculation of
$E(B-V)_{{\rm H}{\alpha}/{\rm H}{\beta}}$ between the previous work
and ours is that we can correct the underlying stellar absorption ${\rm
EW}_{abs}$ from the results of stellar population synthesis without
making any hypotheses. The result of $E(B-V)$ is listed in the last 3 columns of
Table~\ref{tab6}.

From this table, we can find that the internal reddening of the stellar
continuum in BCGs is generally lower than that of ionized gas.	A model
of foreground dust clumps, with different covering factors for gas and
stars, is a possible explanation for the difference. The covering factor
by dusty clumps is greater for the gas region that generates the emission
lines than for the stars that produce the continuum. That the continuum
emission of stars is less obscured than are the emission lines of ionized
gas, has been pointed out for other kinds of emission line galaxies
(Calzetti et al. \cite{calzetti94}).


\section{Summary}

We have observed the optical spectra for the nuclear regions of 10 blue
compact galaxies. We have studied their stellar populations by matching
the spectra of these objects to a library of integrated spectra of star
clusters. Our conclusions can be summarized as follows.

The quantitative analysis indicates that the nucleus of the 3 bright
BCGs is dominated by young components and the star-forming process
is still ongoing. The maximum metallicity of the stellar population
is $[Z/Z_{\odot}]=-0.5$.  The nucleus of the dwarf BCGs (BCDGs, which have
$M_{\rm B}>-20$), on the other hand, is dominated by the intermediate
age component. The metallicity has, at most, reached up to the solar
value. The young component is not so important as in the bright BCGs,
but it is still not negligible.

For all BCGs, the old population in the range $[Z/Z_{\odot}] \leq 0.0$
is important, the very metal-rich component provides quite a small
contribution in these galaxies. The stellar populations of BCGs suggest
that they are old galaxies with intermittent star formation history.

A good match can be achieved between the synthesized and observed spectrum
of BCGs. It suggests that the stellar radiation is an important energy
source for BCGs.

The emission-line spectra from the gaseous components in these objects
were isolated and analyzed. Using these stellar subtracted spectra, we
have calculated the internal reddening value of the emission line regions
and attempted to identify the ionizing mechanism in BCGs. Comparing with
the reddening derived from the continuum, we conclude that the continuum
and emission line regions have different degrees of dust obscurations.

The stellar subtracted spectra should be very useful for further
investigation of physical conditions and chemical abundance of the
emission line regions of BCGs.

\begin{acknowledgements} 
We are grateful to the Chinese 2.16m Telescope
time allocation committee for their support of this programme and to
the staff and telescope operators of the Xinglong Station of Beijing
Astronomical Observatory for their support.  Especially we would like
to thank Prof. J.Y. Hu and Dr. J.Y. Wei for their active cooperation
which enable all of the observations to go through smoothly. We are
also deeply grateful to Henrique R. Schmitt for kindly providing us the
procedure of stellar population synthesis.  We also thank the anonymous
referee for helpful comments and constructive suggestions. Special Thanks
to Dr. S. Mao and Dr. T. Kiang for their hard work of English revision of this paper.
This work was supported by grants from	the National Panden Project and
Natural Science Foundation of China.  
\end{acknowledgements}

\end{document}